\documentclass[aps,prl,twocolumn,superscriptaddress]{revtex4-2}
\usepackage[pdftex]{graphicx}
\usepackage{textgreek}
\usepackage{amssymb}
\usepackage{amsmath}
\usepackage{longtable}
\usepackage{xspace}
\usepackage[hidelinks]{hyperref}
\usepackage{orcidlink}
\usepackage{siunitx}
\newcommand{\Yb}{\textsuperscript{171}Yb\textsuperscript{+}\xspace}

\begin{document}

\title{International atomic clock comparison achieving $\mathbf{5\times 10^{-18}}$ uncertainty}
\newcommand{\PTB}{%
  Physikalisch-Technische Bundesanstalt,
  Bundesallee 100, 38116 Braunschweig, Germany}

\newcommand{\NPL}{%
  National Physical Laboratory,
  Hampton Road, Teddington, TW11 0LW, United Kingdom}

\newcommand{\LUH}{Institut für Erdmessung, Leibniz Universit\"at Hannover (LUH), Schneiderberg 50, 30167, Hannover, Germany}

\newcommand{\LTE}{Laboratoire Temps Espace (LNE-OP), Observatoire de Paris, Université PSL, Sorbonne Université, Université de Lille, LNE, CNRS, 61 avenue de l'Observatoire, 75014 Paris, France}

\newcommand{\LPL}{Laboratoire de Physique des Lasers (LPL), Université Sorbonne Paris Nord, CNRS, 99 Avenue Jean-Baptiste Clément, 93430 Villetaneuse, France.}

\author{A.~Tofful\,\orcidlink{0000-0001-5907-6657}}
\thanks{These authors contributed equally to this work.}
\affiliation{\NPL}

\author{M.~Filzinger\,\orcidlink{0000-0001-5612-2240}}
\thanks{These authors contributed equally to this work.}

\author{T.~Schmidt\,\orcidlink{0009-0003-5592-3582}}
\thanks{These authors contributed equally to this work.}

\affiliation{\PTB}
\author{A.~Amy-Klein\,\orcidlink{0000-0002-5122-7868}}
\affiliation{\LPL}

\author{E.~Benkler\,\orcidlink{0000-0001-7907-849X}}
\affiliation{\PTB}

\author{E.~Cantin\,\orcidlink{0000-0001-9362-0628}}
\affiliation{\LPL}

\author{C.~Chardonnet\,\orcidlink{0009-0007-3194-8119}}
\affiliation{\LPL}

\author{E.~A.~Curtis\,\orcidlink{0000-0003-3487-5588}}
\affiliation{\NPL}

\author{H.~Denker\,\orcidlink{0000-0001-9112-1729}}
\affiliation{\LUH}

\author{B.~Dutta\,\orcidlink{0009-0005-7719-4012}}
\affiliation{\LPL}

\author{J.-O.~Gaudron}
\affiliation{\NPL}

\author{W.~Huang\,\orcidlink{0000-0002-9998-9730}}
\affiliation{\NPL}

\author{J.~Kronjäger\,\orcidlink{0000-0002-5181-1407}}
\affiliation{\PTB}

\author{A.~Kuhl\,\orcidlink{0000-0003-4306-8075}}
\affiliation{\PTB}

\author{R.~Le~Targat\,\orcidlink{0000-0002-8740-1709}}
\affiliation{\LTE}

\author{B.~Lipphardt\,\orcidlink{0000-0001-7039-1914}}
\affiliation{\PTB}

\author{O.~Lopez\,\orcidlink{0000-0003-3171-2863}}
\affiliation{\LPL}

\author{S.~Mukherjee\,\orcidlink{0000-0003-4857-6423}}
\affiliation{\PTB}

\author{A.~L.~Parke\,\orcidlink{0009-0001-8138-6085}}
\affiliation{\NPL}

\author{B.~Pointard\,\orcidlink{0000-0003-4503-3546}}
\affiliation{\LTE}

\author{P.-E.~Pottie\,\orcidlink{0000-0003-3677-2208}}
\affiliation{\LTE}

\author{P.~J.~Regan\,\orcidlink{0009-0008-3241-8344}}
\affiliation{\NPL}

\author{M.~Risaro\,\orcidlink{0000-0001-8046-5710}}
\affiliation{\NPL}

\author{M.~Schioppo\,\orcidlink{0000-0002-0197-2763}}
\affiliation{\NPL}

\author{M.~Steinel\,\orcidlink{0000-0002-4876-083X}}
\affiliation{\PTB}

\author{J.~Tunesi\,\orcidlink{0000-0003-1228-674X}}
\affiliation{\NPL}

\author{X.~Zhang\,\orcidlink{0000-0002-7652-4797}}
\affiliation{\NPL}

\author{R.~M.~Godun\,\orcidlink{0000-0001-5256-9088}}
\email{rachel.godun@npl.co.uk}
\affiliation{\NPL}

\author{N.~Huntemann\,\orcidlink{0000-0002-1876-0513}}
\email{nils.huntemann@ptb.de}
\affiliation{\PTB}

\date{\today}

\begin{abstract}

Two single-ion optical clocks, located in the United Kingdom and in Germany, are compared remotely via a 2220\,km long optical fiber link. Both clocks are based on the ${}^2S_{1/2} (F=0)\leftrightarrow {}^2F_{7/2} (F=3)$ electric octupole (E3) transition in $^{171}\text{Yb}^+$. Using more than $10^6$ seconds of measurement data, we determine a fractional frequency difference of $1.0(4.8)\times10^{-18}$, demonstrating agreement of the independently developed systems. To our knowledge, this constitutes the first international clock comparison reaching a fractional uncertainty below $5\times10^{-18}$, an important milestone towards an anticipated optical redefinition of the SI second. The measurement uncertainty is dominated by that of the relativistic redshift correction between the two sites, which includes time-varying gravity potential differences.
We exploit the large baseline between the two clocks to search for violations of Lorentz invariance. In the absence of statistically significant modulations, we improve constraints on the Robertson–Mansouri–Sexl parameter $\alpha$ by approximately a factor of five in the uncertainty, and provide limits on corresponding parameters within the Standard-Model Extension framework.
\end{abstract}

\maketitle

Optical atomic clocks, which now feature fractional frequency uncertainties at and even below the $10^{-18}$ level, have applications in fundamental physics, geodesy, and timekeeping~\cite{Ludlow2015}. As they outperform by more than two orders of magnitude the best cesium microwave clocks serving as the primary realizations of the SI second, a future optical redefinition of the second is anticipated~\cite{Dimarcq2024}.

Working towards selecting a new primary standard for time and frequency, the performance of optical clocks must be verified through comparisons. A few such frequency comparisons between clocks based on different atomic reference transitions have already been performed with mid-$10^{-18}$ uncertainties \cite{Hausser2025, Aeppli2026, Filzinger2026}. The resulting non-unity frequency ratios can be compared to those determined elsewhere, but the unperturbed values are not known a priori, making it harder to identify and resolve discrepancies, such as those reported in~\cite{Aeppli2026, Doerscher2021}. In contrast, comparisons of systems based on the same reference transition should provide a frequency ratio of one, which is a clear benchmark. Although this type of comparison has been performed with even smaller uncertainties~\cite{McGrew2018,Sanner2019a, Zhang2023,Arnold2026, Pizzocaro2026, Vishwakarma2026} than obtained with different species clocks, agreement  with $< 1 \times 10^{-17}$ uncertainty has almost always been observed between systems developed and operated within a single institute. The only exception so far has been the agreement of the same two systems as described in this work as part of an earlier measurement campaign between several European institutes \cite{Pizzocaro2026}. The obtained uncertainty, however, was limited by statistics and did not reach the target level of $5 \times 10^{-18}$, required ahead of redefining the second~\cite{Dimarcq2024}.
Comparisons between other independently developed same-species clocks have revealed discrepancies well beyond the stated measurement uncertainties, indicating the presence of unaccounted-for systematic effects~\cite{ICON2024, Lindvall2025}. 

In this Letter, we report on a comparison between two independently developed optical clocks operated in different countries that show better agreement than the target level of $5 \times 10^{-18}$.
Both clocks use the ${}^2S_{1/2} (F=0)\leftrightarrow {}^2F_{7/2} (F=3)$ electric octupole (E3) transition of a single trapped ytterbium ion \Yb as the reference. One clock, NPL-E3Yb+3, is developed and operated at the National Physical Laboratory (NPL) in the United Kingdom, while the other, PTB-Yb1E3, is operated at the Physikalisch-Technische Bundesanstalt (PTB) in Germany. We find a frequency difference between the two clocks of $1.0(4.8)\times10^{-18}$, where the relative standard uncertainty is dominated by that of the relativistic redshift between the distant clocks. We also use the measurement data to search for violations of Lorentz invariance. In the absence of statistically significant modulations, we improve limits on the Robertson-Mansouri-Sexl (RMS) parameter $\alpha$~\cite{Delva2017} by a factor of almost five in the uncertainty and obtain corresponding parameters within the Standard-Model Extension (SME) framework~\cite{Qin2025}.

\begin{figure}
\includegraphics[width=.8\linewidth]{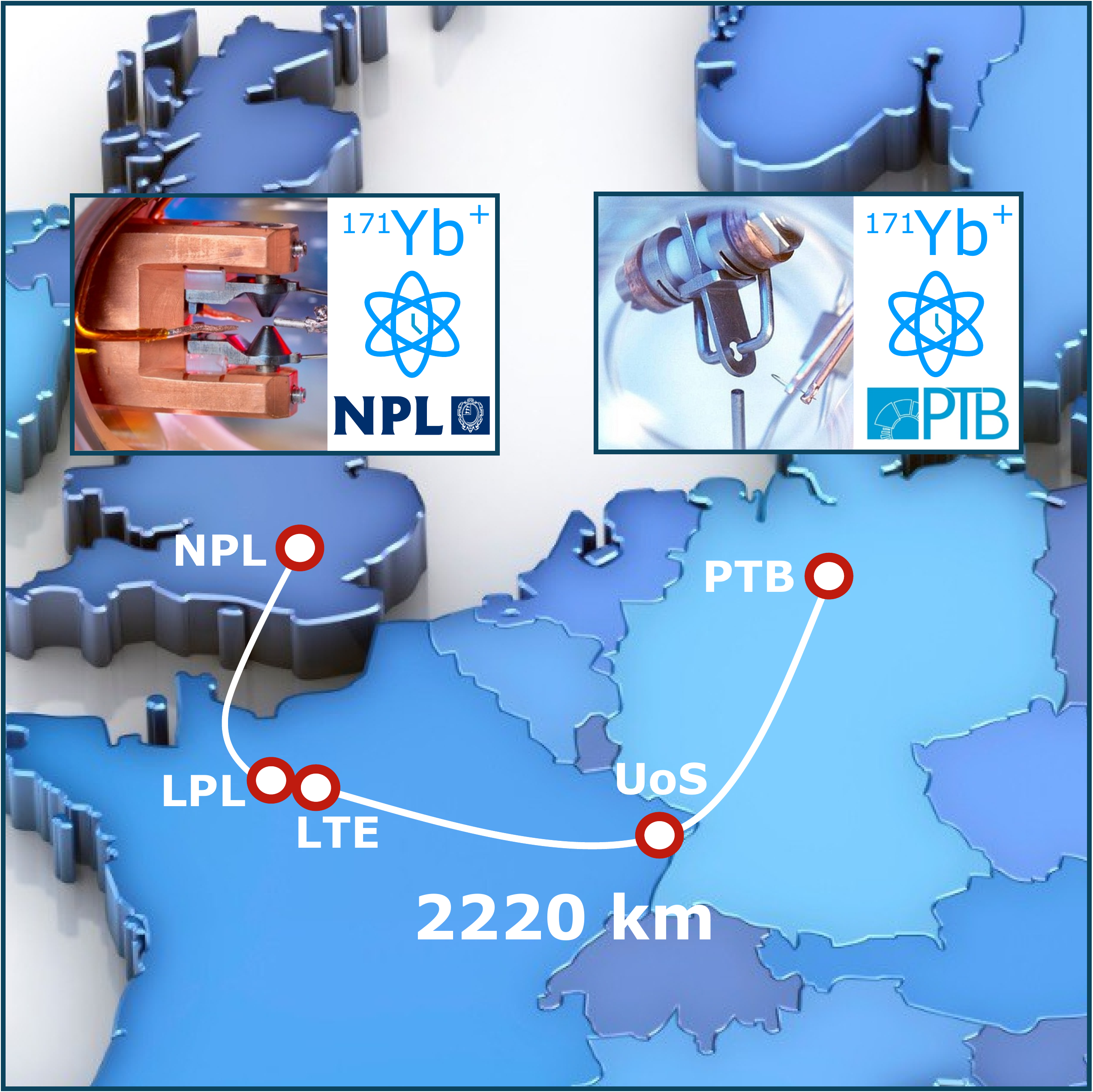}
\caption{Map indicating the locations of the two optical clocks at NPL, UK, and PTB, Germany, and their connection via a 2220\,km long fiber link with three nodes in France at Laboratoire de Physique des Lasers (LPL), Laboratoire Temps Espace (LTE), and the University of Strasbourg (UoS). Insets show the different geometries of the respective single-ion trap electrodes. \label{fig:map}}
\end{figure}

NPL-E3Yb+3 and PTB-Yb1E3 were compared between 18 March and 12 April 2025. The two clocks are connected within the European network of phase-stabilized fiber links~\cite{Schioppo2022}, on a span of 2220\,km via three nodes in France at LPL, LTE and University of Strasbourg, as illustrated in~\autoref{fig:map}. 

For each optical clock, a single \Yb ion is confined in a radiofrequency (RF) trap. The system at PTB uses conventional Paul-trap geometry with a ring electrode, while the trap at NPL is based on an end-cap trap~\cite{NisbetJones2016} (see insets in~\autoref{fig:map}). In both cases, clock operation is performed by repeating an experimental sequence involving laser cooling, state preparation, probing of the E3 transition, state readout, and frequency steering. Laser cooling and state detection are performed on the dipole-allowed $^2S_{1/2}\leftrightarrow {}^2P_{1/2}$ transition at 370\,nm, aided by repumping from the metastable $^2D_{3/2}$ state with a laser at 935\,nm. To overcome coherent population trapping at low magnetic fields~\cite{Ejtemaee2010}, a magnetic field of a few hundred microtesla is applied during cooling and detection. The ion is subsequently prepared in the $F=0$ ground state and interrogated on the E3 transition. The transition is probed at a small magnetic field strength of about 3~\textmu T. 
%Decay of the larger magnetic field before clock interrogation adds about 50\,ms and 25\,ms of dead time to the clock sequences at PTB and NPL respectively.
To enable sufficient decay of the larger magnetic field before clock interrogation, 50\,ms and 25\,ms of dead time are added to the clock sequences at PTB and NPL respectively.
After interrogation, the absence of fluorescence in the presence of cooling laser light indicates excitation of the clock transition. Repumping from the $^2F_{7/2}$ state is performed with laser light at 760\,nm. 

At both institutes, clock laser light at 467\,nm is obtained via frequency doubling from 934\,nm, and narrow sub-Hz linewidths are achieved via reference to an ultrastable cavity. At NPL, the 934\,nm laser is referenced to a 1064\,nm laser through an optical frequency comb. The 1064\,nm laser is locked to a room-temperature ultrastable cavity based on a 48.5\,cm long ULE spacer, providing the high frequency stability required for the comparison~\cite{Schioppo2022}. At PTB, the 934\,nm laser is stabilized via a frequency comb to a 1542\,nm laser, which in turn is locked to an ultrastable cryogenic silicon cavity~\cite{Matei2017}.

The two optical clocks differ significantly in their implementations. Notably, the clocks at NPL and PTB used different methods for light shift cancellation: due to the multiyear lifetime of the \Yb E3 transition's excited state~\cite{Lange2021,Lange2021E} and corresponding small coupling strength, laser excitation of this transition leads to an exceptionally large light shift. Several methods have been developed over the years to control this shift~\cite{Yudin2010,Huntemann2012a, Huntemann2012, Sanner2018, Lange2022}. At NPL, the E3 transition was probed in an interleaved dual-servo fashion with Rabi $\pi$-pulses at alternating powers. This enabled extrapolation to a light-shift-free frequency at zero optical power. A probe power ratio of $\sim$10:1 was used, with $\pi$-pulse durations of 290\,ms and 900\,ms, resulting in light shifts of $\sim$6.3\,Hz and $\sim$0.6\,Hz, respectively. At PTB, the transition was obtained using Rabi-controlled hyper-Ramsey spectroscopy (HRS) ~\cite{Huntemann2016} with a Ramsey dark time of 500\,ms. 
%Here, the result of each hyper-Ramsey interrogation is light-shift free, while an additional servo using Rabi interrogation with $60$\,ms long pulses is used to track variations of the light shift, ensuring resonant Ramsey pulses. 
Here, the probe-induced light shift is coherently suppressed within each HRS interrogation cycle. Residual variations of the light shift are tracked by an additional servo loop based on Rabi interrogation with $60\,\mathrm{ms}$ pulses. The resulting estimate of the light shift is used to adjust the frequency step applied during the HRS interrogation pulses, thereby maintaining cancellation of the probe-induced light shift. The associated uncertainty arises from possible differences between the Rabi and HRS interrogation conditions, as well as from the statistical uncertainty of the control loop.
Additional details on the experimental setup and the mode of operation of NPL-E3Yb+3 and PTB-Yb1E3 can be found respectively in Ref.~\cite{Tofful2024} and Refs.~\cite{Sanner2019a, Huntemann2016}.

Table \ref{tab:combined_budget} reports the systematic frequency shifts and their uncertainties for both clocks.  
In each case, the largest shift of the transition frequency was the Stark shift caused by black-body radiation (BBR) from the ion’s room-temperature environment. Both evaluations of this shift make use of the same differential polarizability coefficient~\cite{Huntemann2016}, leading to a correlated uncertainty. 
We therefore split the evaluation of the BBR shift and report it as a common shift at a reference temperature of $294.62$\,K and an offset that depends on the BBR temperature in each apparatus. At NPL, the reference temperature is realized with $0.15$\,K uncertainty, while at PTB the BBR temperature is larger by 4.8(1.3)\,K. This difference results from the higher room temperature and the larger temperature rise of the ion trap during operation. 

The remaining frequency shifts are treated as uncorrelated between the two clocks because the uncertainties of common atomic parameters cause differences smaller than $10^{-19}$ in the corresponding differential uncertainties. 

In both systems, the electric quadrupole shift was evaluated on the ${}^2S_{1/2} (F=0)\leftrightarrow \,{}^2D_{3/2} (F=2)$ electric quadrupole (E2) transition and scaled down by $-92.1(1.7)$ for the E3 transition~\cite{Lange2020}. The shift was measured by interrogating the E2 transition in three orthogonal magnetic field directions, including the field direction used for probing the E3 transition. For NPL-E3Yb+3, this method differs from the previous evaluation~\cite{Tofful2024} where the E2 transition was probed in five different magnetic field directions and therefore needed longer averaging times to reach the same level of uncertainty. The shift was evaluated for NPL-E3Yb+3 immediately before and after the clock comparison campaign, with the two results agreeing within their combined uncertainty. For PTB-Yb1E3, the shift was instead measured during clock operation with interleaved interrogations of the E2 transition~\cite{Filzinger2023}.

The leading contribution to the total measurement uncertainty comes from correcting for the differential relativistic redshift between the two clocks in their respective gravity potentials in the UK and Germany. In addition to correcting for the mean (static) gravity potential difference between the clocks~\cite{Riedel_2020}, we also applied a further (dynamic) correction to take account of the differential tidal potential sampled by the clocks~\cite{Voigt2016} during the measurement campaign. The total uncertainty is equivalent to a relative geopotential-height uncertainty between the clocks of just over 3\,cm.

\begin{table*}
\sisetup{table-alignment-mode = marker}
\caption{Leading frequency shifts and corresponding uncertainties of the ${}^2S_{1/2} (F=0)\leftrightarrow {}^2F_{7/2} (F=3)$ clock transition frequency of NPL-E3Yb+3 and PTB-Yb1E3 during the period between 18 March and 12 April 2025. The fractional shifts and associated 1-sigma uncertainties are reported in units of $10^{-18}$ and rounded to the $10^{-19}$ level.}
\label{tab:combined_budget}
\begin{ruledtabular}
\begin{tabular}{lSSSSS}
\textbf{Systematic effect}   & \textbf{$\Delta\nu_{\text{NPL}}/\nu_0$} & \textbf{$\sigma_{\text{NPL}}/\nu_0$} & \textbf{$\Delta\nu_{\text{PTB}}/\nu_0$} & \textbf{$\sigma_{\text{PTB}}/\nu_0$} &
\textbf{$\sigma_{\text{differential}}/\nu_0$}\\
\hline
Black-body radiation (@294.62~K)& -67.1 & 1.2 & -67.1 & 1.2 & 0\\
Black-body radiation (residual)& {0\,\,\textsuperscript{a}} & 0.2 & -4.5 & 1.3 & 1.3
\\
Background gas collisions & 0 & 0.6 & 0 & 0.5 & 0.8\\
Electric quadrupole & 22.5 & 0.6 & -4.4 & 0.5 & 0.8\\
Quadratic Zeeman (DC) & -29.0 & 0.6 & -28.8 & 0.2 & 0.6\\
Second-order Doppler & -1.7 & 0.4 & -2.3 & 1.5 & 1.6\\
Phase chirp & 0 & 0.4 & 0 & 0.1 & 0.4\\
AC Stark - probe beam & {0\,\,\textsuperscript{a}} & 0.2 & 0 & 0.8 & 0.8\\
Servo offset & 0 & 0.1 & 0 & 0.2 & 0.2\\
Trap-induced AC Zeeman & 0.2 & 0.1 & 0.6 & 0.6 & 0.6\\ 
Trapping RF Stark & -0.4 & 0.1 & -0.8 & 0.6 & 0.6\\
\hline
\textbf{Total (clocks)} & -75.5 & 1.8 & -107.3 & 2.7 & 2.8 \\
\hline
Relativistic redshift (static+dynamic) & {1186.9+8.7\,\,\textsuperscript{a}} & 2.5 & {8487.4+9.4\,\,\textsuperscript{a}} & 2.4 & 3.5\\ 
%Relativistic redshift () & $8.7$ & $0^{b}$ & $9.4$ & $0^{b}$ & $0^{b}$\\ 
Comb and local links & 0 & 0.4 & 0 & {$<0.1$} & 0.4\\
International fiber link & {$-$} & {$-$} & {$-$} & {$-$} & 0.9\\ 
\hline
\textbf{Total} & 1120.2 & 3.1 & 8389.5 & 3.6 & 4.6\\
\end{tabular}
\begin{flushleft}
    $^{\text{a}}$ These shifts were dynamically corrected (either in post-processing or in real time).
    
\end{flushleft}
\end{ruledtabular}
\end{table*}

\begin{figure}[h!]
    \centering
    \includegraphics[width=\linewidth]{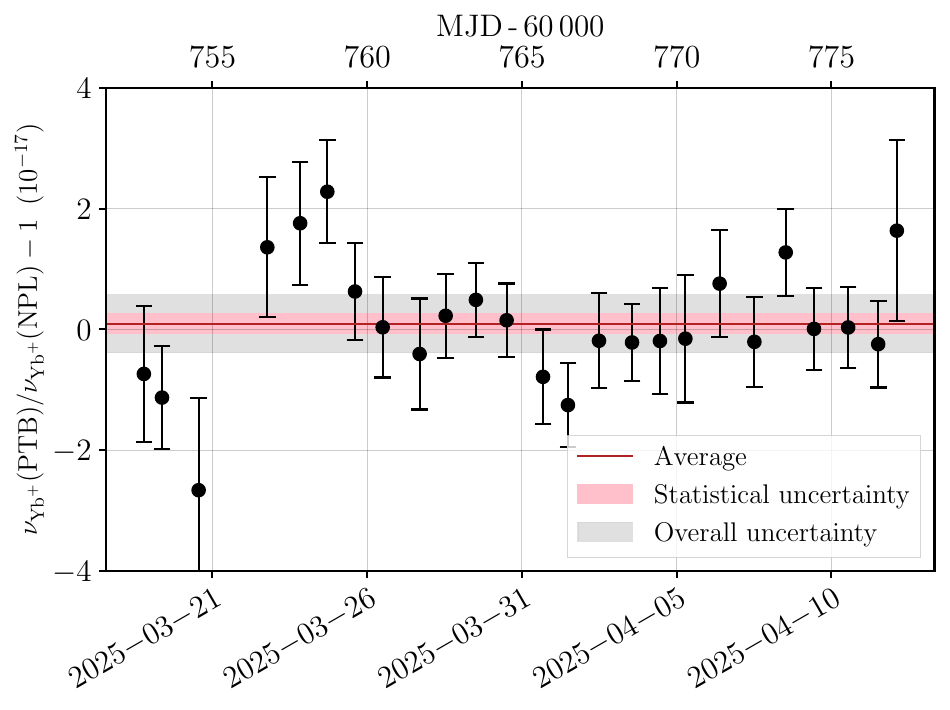}
    \caption{Frequency ratio measurements between $\nu_{\text{Yb}^+}$ of the ${}^2S_{1/2} (F=0)\leftrightarrow {}^2F_{7/2} (F=3)$ transition in \Yb as realized independently at PTB and NPL. Each point represents an average over an individual day, and error bars correspond to the one-sigma statistical uncertainties, assuming white frequency noise. We find $\chi^2_\textrm{red} \approx 1.2$. The red line corresponds to the average of all measurements, the pink shaded area depicts the statistical uncertainty of $1.6\times 10^{-18}$ and the grey shaded area corresponds to the overall uncertainty of $4.8\times 10^{-18}$.}
    \label{fig:daily_ratio}
\end{figure}

Both at PTB and NPL, a single branch of a frequency comb was used for measuring the optical frequency ratio between the clock and a 1542 nm transfer laser used in the international fiber link~\cite{Benkler2019}. In each case, optical path-length stabilization ensures phase coherent measurements. At PTB, this leads to a negligible contribution to the systematic uncertainty, below $10^{-19}$ for the entire local measurement, while at NPL the uncertainty is currently limited to $3.8\times 10^{-19}$.

Larger uncertainties arise from the fiber link between NPL and PTB operated in segments between the locations displayed in~\autoref{fig:map}. While for the links between PTB and UoS \cite{Raupach2015,Koke2019} and UoS and LPL \cite{Cantin_2021} an uncertainty below $10^{-19}$ is achieved, the link between NPL and LPL dominates the total link uncertainty~\cite{Schioppo2022}. For the given measurement period, its systematic uncertainty was $9\times10^{-19}$.

We evaluated $1.09\times10^6\,$s (12.6 days) of measurement data taken within a period of 24.7 days between March 18, 2025 (MJD 60\,752) and April 12, 2025 (MJD 60\,777). 
%[Something about the overall uptime as well as those of each clock, and the limiting factors]
During this time, the experimental availabilities of the clocks at NPL and PTB were 92\% and 86\%, respectively, while the combined uptime including the link was 51\% (see Supplemental Material for details).

\begin{figure}[h!]

    \centering
    \includegraphics[width=\linewidth]{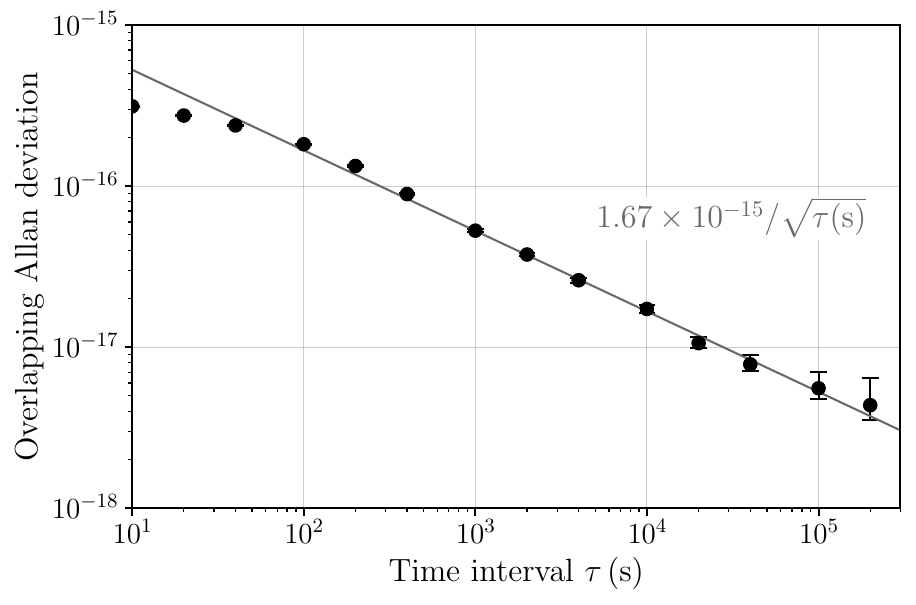}
    \caption{Measurement instability as characterized by the overlapping Allan deviation. The solid line indicates the $1/\sqrt{\tau}$-scaling expected for white frequency noise. An instability of $1.67\times10^{-15}/\sqrt{\tau(\text{s})}$ is obtained for $\tau > 1000\,$s, where quantum projection noise dominates.} %Extrapolation to the full measurement time yields a statistical uncertainty of $u_\text{stat}=1.60\times10^{-18}$.}
    \label{fig:adev}
\end{figure}

The dataset, averaged over individual days, is displayed in~\autoref{fig:daily_ratio}. We find an unperturbed frequency ratio between the E3 transition frequency as realized at PTB and at NPL of 
\begin{equation}
   \frac{\nu_{\text{Yb}^+}\text{(PTB)}}{\nu_{\text{Yb}^+}\text{(NPL)}}= 1.000\,000\,000\,000\,000\,001\,0 (48)\,,
\end{equation}
corresponding to a frequency difference of $1.0(4.8)\times 10^{-18}$ between the two clocks, where the uncertainty is the sum in quadrature of the systematic and statistical uncertainties.
The instability of the measurement, as characterized by the Allan deviation, is shown in~\autoref{fig:adev}. Based on the observed white frequency noise scaling of $1.67\times10^{-15}/\sqrt{\tau(\mathrm{s})}$, the statistical uncertainty for the full dataset is evaluated to be $1.6\times10^{-18}$. The statistical properties of the data are examined in more detail in the Supplemental Material.

The large spatial separation between PTB (52.30°\,N, 10.46°\,E) and NPL (51.43°\,N, 0.34°\,W) makes the comparison sensitive to violations of Lorentz invariance, which would cause the frequency ratio to depend on the clocks' position and velocity relative to a hypothetical preferred frame. We search for such violations in two complementary approaches.
In the Robertson--Mansouri--Sexl (RMS) formalism~\cite{Robertson1949,MansouriSexl1977}, an anisotropy of the speed of light is parametrized by a single dimensionless coefficient $\alpha$, which vanishes within special relativity. As Earth's rotation continuously changes the orientation and velocity of the baseline between the two clocks, such an anisotropy would produce a signal in the frequency ratio of the form
\begin{equation}
y_\text{RMS}(t) = \alpha\, Q(t)/c^2,
\label{Eq2}
\end{equation}
oscillating at the sidereal period, where $Q(t)=2\mathbf{w}\cdot\Delta\mathbf{v}(t)$ projects the Earth's velocity $\mathbf{w}$ with respect to the cosmic microwave background (CMB) rest frame onto the differential velocity $\Delta\mathbf{v}(t)$ of the two sites. For PTB--NPL the geometric sensitivity, i.e. the peak value of $Q(t)/c^2$ over a sidereal day, is $4.4\times 10^{-10}$. We fit the data with the expected signal given by Eq.~(\ref{Eq2}) plus a constant offset, with the offset and $\alpha$ as the only free parameters, yielding $\alpha = -2.4(2.3)\times 10^{-9}$, which corresponds to a frequency-ratio modulation amplitude of $-1.0(1.0)\times 10^{-18}$. This improves on the previous best RMS bound of $|\alpha| < 1.1\times10^{-8}$~\cite{Delva2017} by a factor of almost 5 in the uncertainty.
We note that neglecting the dynamic correction for tidal effects between the remote clocks leads to a non-zero periodic variation in the frequency ratio at the sidereal period, at the 2.6-sigma level.

Within the Standard-Model Extension (SME) framework~\cite{Kostelecky2011}, the Sun's gravitational potential may couple anomalously to individual particle species, in contrast to the universal coupling of general relativity. Under a universal coupling, the leading first-order solar term is unobservable in a comparison of two Earth-fixed clocks. It corresponds to a uniform gravitational acceleration across the Earth, which the Earth's free fall cancels exactly, leaving only the second-order tidal variations corrected for above. A particle-dependent coupling would impair this cancellation and leave a residual first-order shift in the frequency ratio of the form
\begin{equation}
y_\text{SME}(t) = (\xi-\beta)\,\Delta U_S(t)/c^2\, ,
\label{Eq3}
\end{equation}
oscillating at the period of a solar day. Here, $\Delta U_S(t)$ is the first-order difference of the Sun's potential between the two sites and $(\xi-\beta)$ the particle-dependent coupling that survives the free fall. Being first-order in the solar field, this signal is dominated by the Sun, with the lunar contribution below 1\%, in contrast to the second-order tidal term, which the Moon dominates. The geometric sensitivity, i.e.~the peak value of $\Delta U_S(t)/c^2$ over a solar day, is $5.0\times10^{-14}$. Fitting the expected dependence given by Eq.~(\ref{Eq3}), together with a free constant offset, yields a coupling $(\xi-\beta) = -2.4 (2.0)\times 10^{-5}$, corresponding to a frequency-ratio modulation amplitude of $-1.2 (1.0)\times 10^{-18}$.
To place bounds on individual SME Lorentz-violating coefficients $\bar{c}_{TT}^w$ ($w=e,p,n$), which quantify the coupling of electrons, protons, and neutrons to a preferred frame, we use a maximum-reach analysis, constraining one coefficient at a time and setting the other coefficients to zero. The sensitivity of the \Yb E3 transition to the electron sector is given by the relativistic factor $R=-1.9$~\cite{Dzuba2017}, obtained from ab initio atomic structure calculations. No analogous calculations exist for the proton and neutron sectors, where all existing bounds rely on the hydrogenic approximation~\cite{Kostelecky2011, Qin2025}. Here, the sensitivities are obtained from nuclear mass fractions alone and nuclear structure effects are neglected. With these sensitivities, our data yield the following bounds at $1\sigma$:
\begin{align}
    \bar{c}_{TT}^{e} &= -1.9 (1.5) \times 10^{-5} \notag\\
    \bar{c}_{TT}^{p} &= -3.9 (3.2) \times 10^{-5} \\
    \bar{c}_{TT}^{n} &= -3.3 (2.7) \times 10^{-5} \notag.
\end{align}
Among comparisons of identical clocks at remote locations, our bounds improve on the previous best limits~\cite{Qin2025} by a factor of $\sim$4.5. However, comparisons of spectrally distinct transitions, which feature a large differential sensitivity $\Delta R$ between them, within the same laboratory already constrain the nuclear sectors at the $10^{-7}$ level and the electron sector at the $10^{-8}$-$10^{-9}$ level~\cite{Qin2025,Kostelecky2011b}.

To summarize, a remote comparison of two independently developed clocks based on the same electric octupole transition in \Yb has yielded agreement at an uncertainty below $5\times10^{-18}$, an important step towards the redefinition of the second. The measurement was performed via a 2220\,km long fiber link. The large separation of the two clocks provides sensitivity to possible violations of Lorentz invariance. In the absence of statistically significant signals, we provide improved constraints. 

Since the relativistic redshift correction between the two sites contributed the largest uncertainty to the measurement, performing even more accurate international clock comparisons requires improved knowledge of the gravity potential. Alternatively, agreement between two distant clocks could be tested if one of them could be transported to the remote location without sacrificing performance. 

\subsection*{Acknowledgments}
We thank Ian Hill for his careful reading of the manuscript and for many valuable suggestions.
We thank Uwe Sterr and Thomas Legero for providing a laser stabilized to an ultrastable silicon resonator. 
We acknowledge support from the UK Department for Science, Innovation and Technology through the National Measurement System programme and the projects 22IEM01 TOCK and 23FUN03 HIOC.
The projects (22IEM01 TOCK, 23FUN03) have received funding from the European Partnership on Metrology, co-financed from the European Union’s Horizon Europe Research and Innovation Programme and by the Participating States.
PTB acknowledges support from the German Research Foundation (DFG) under SFB~1227 DQ-mat -- Project-ID 274200144 -- within project B02 (320087752);  and by the Max Planck–RIKEN–PTB Center for Time, Constants and Fundamental Symmetries.
Optical fiber links from the French REFIMEVE network are supported by Program “Investissements d’Avenir” launched by the French Government and implemented by Agence Nationale de la Recherche with reference ANR-21-ESRE-0029(ESR/Equipex+T-REFIMEVE).

\newpage
\subsection*{Supplemental Material}

\begin{figure*}[t]
    \centering
    \includegraphics[width=\linewidth]{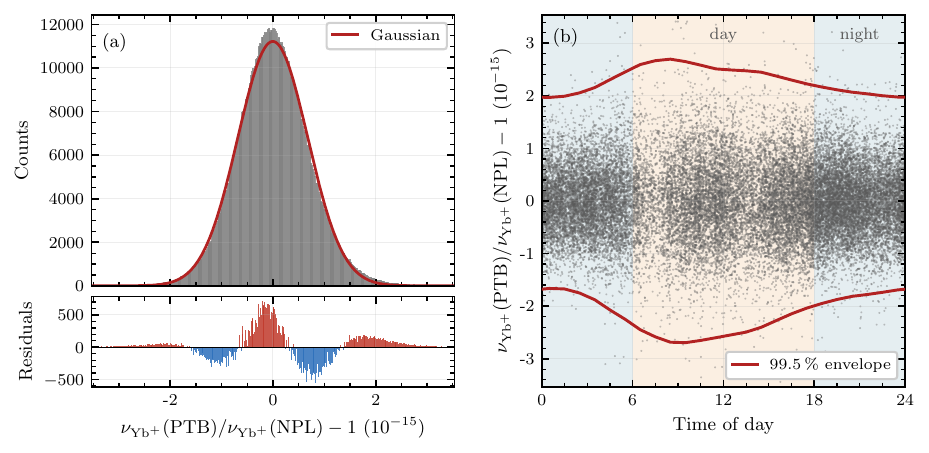}
    \caption{(a) Distribution of the 1-second frequency-ratio samples (gray bars) with a Gaussian fit (red line); the lower panel shows the residuals (data $-$ fit), red and blue marking bins where the data lie above and below the fit. (b) The same samples versus time of day (a random subsample of 30\,000 of the $1.09\times 10^6$ samples is shown), with the central 99.5\,\% envelope per hour (red); the shaded regions mark daytime (06:00--18:00 UTC) and nighttime. The distribution is broader during the day than at night by a factor of 1.24.\label{fig:dist}}
\end{figure*}

We examined the distribution of the 1-second frequency-ratio samples $R_i$ shown in \autoref{fig:dist}(a), with sample mean $\bar{R}$ and standard deviation $\sigma$. We quantify its shape by the skewness $\gamma_1 = \frac{1}{N}\sum_i(R_i-\bar{R})^3/\sigma^3$ and the excess kurtosis $\gamma_2 = \frac{1}{N}\sum_i(R_i-\bar{R})^4/\sigma^4 - 3$. Both vanish for a normal distribution, and a positive $\gamma_2$ indicates heavier tails with a sharper central peak. For our data we find $\gamma_1 = 0.09(2)$ and $\gamma_2 = 1.36(20)$, indicating a slight asymmetry and heavier tails than expected for a Gaussian distribution.

In contrast with this dataset, local frequency ratio measurements between co-located clocks in both institutes were observed to be well described by a Gaussian distribution. Furthermore, the distribution of the frequency ratio measured over the fiber link is broader during the daytime than at night by 24~\% (\autoref{fig:dist}(b)), consistent with increased environmental perturbations of the link. Together, these observations suggest that the heavy tails and the skewness are associated predominantly with noise introduced by the fiber link rather than with the clocks themselves. 

\begin{figure*}[t]
    \centering
    \includegraphics[width=\linewidth]{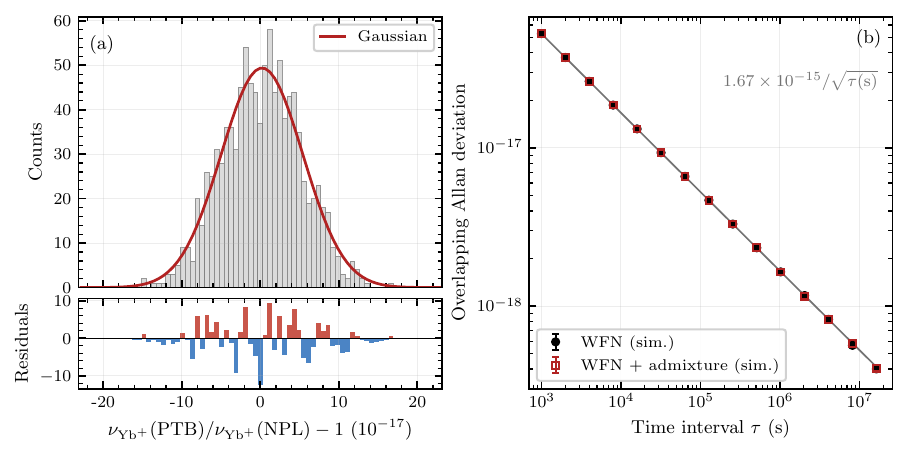}
    \caption{Influence of the non-Gaussian distribution on the statistical uncertainty, shown for the 1000~s averages of the data. (a) Distribution of consecutive 1000~s averages of the measured frequency-ratio samples (gray) with a Gaussian fit (red). The lower panel shows the residuals (data $-$ fit). The averages retain an excess kurtosis of $\gamma_2 = 1.0(7)$, caused by the changes in variance of the fiber-link noise throughout the day. Their skewness of $\gamma_1 = -0.14(18)$ is consistent with zero, and their standard deviation of $5.14(14)\times10^{-17}$ agrees with the white-noise expectation of $1.67\times10^{-15}/\sqrt{1000\,\mathrm{s}} = 5.28\times10^{-17}$. (b) Overlapping Allan deviation of $10^{7}$ simulated uncorrelated 1000~s samples drawn from a distribution matched to the skewness and excess kurtosis of the measured 1000~s averages (WFN + admixture, open squares) and of Gaussian white noise of equal variance (WFN, filled circles). The two agree at all averaging times.\label{fig:sim1000}}
\end{figure*}

In the recorded data, out of a total of $1.09\times10^{6}$ samples, 0.14\,\% deviate by more than $4\sigma$, compared with 0.006\,\% expected for a Gaussian distribution of the same width. 
These samples primarily lead to the nonzero excess kurtosis. 
In the 10\,s before and after each $>3\sigma$ deviation, the data are offset from the campaign mean in the same direction as the deviation itself, by approximately $2\times10^{-16}$ on average. This points to an underlying physical process causing the offsets, rather than statistical anomalies. We attribute these excursions to disturbances on the fiber link. 
Because the optical phase of the link is actively stabilized, the  excursions are typically followed by compensating  offsets over the subsequent minutes. This effect leads to the observed asymmetry: there are more moderately large positive samples, balanced by a larger number of smaller negative ones.
These excursion-recovery sequences leave no net contribution to the mean frequency ratio, and averaged over 1000\,s the asymmetry is negligible: $\gamma_1 = -0.14(18)$ (\autoref{fig:sim1000}). We note that removing outliers based on a symmetric distribution from the 1-second samples could remove only one part of an excursion, which would introduce a bias in the mean frequency ratio, and is therefore avoided.

\begin{figure*}[t]
    \centering
    \includegraphics[width=\linewidth]{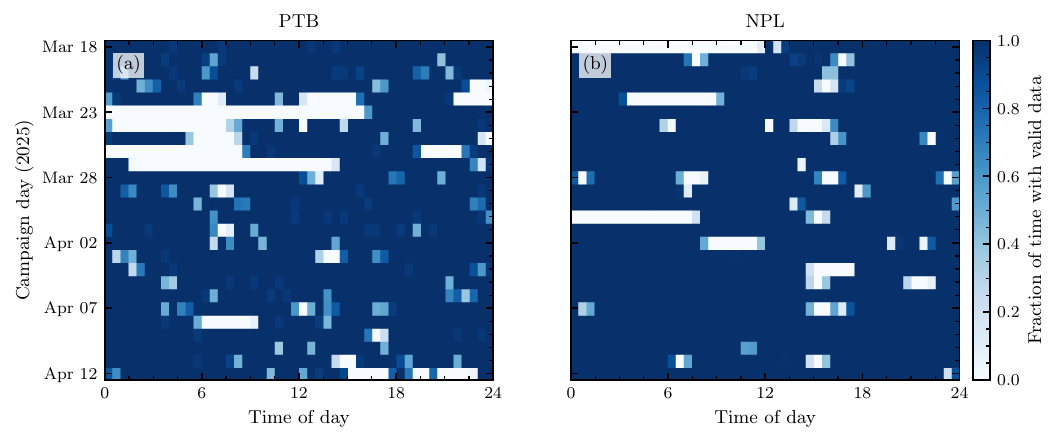}
    \caption{Availability of the two individual clocks: fraction of each 30-minute interval with valid data at (a) PTB and (b) NPL, for each day of the campaign.\label{fig:mask_clocks}}
\end{figure*}

The conservation of the mean likewise requires that these excursion-recovery sequences are recorded in their entirety. Data gaps could violate this if a part of such an event is removed, and we therefore examined whether this was the case. Gaps are present where a subsystem was not operating or a reading was flagged invalid. Long gaps, which account for most of the missing time, are dominated by the availability of the two clocks and their local measurement chains at the two sites (\autoref{fig:mask_clocks}). Short gaps are predominantly associated with local validity flags, e.g. cycle-slip checks of the counters at the frequency combs. In this case, their flagged readings remained recorded so their content can be inspected.
Gaps associated with significant frequency excursions are linked to failure events of the phase or frequency stabilization systems within the measurement infrastructure. These data had been excluded based on validity flags independent of the value of the frequency ratio.  To test whether the remaining gaps introduced a bias, we excluded data surrounding each gap and recomputed the mean frequency ratio. Even when removing data within 120\,s of each gap, the mean changed by only $0.6\times10^{-18}$. This indicates that data gaps do not preferentially remove portions of excursion-recovery sequences and therefore do not introduce a significant bias in the mean frequency ratio.

\begin{figure*}[t]
    \centering
    \includegraphics[width=\linewidth]{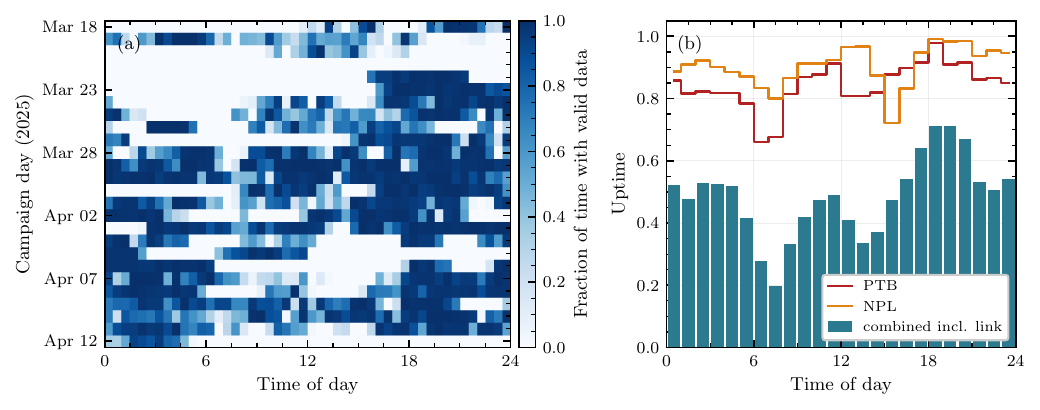}
    \caption{Sampling of the combined frequency-ratio data. (a) Fraction of each 30-minute interval with valid data, for each day of the campaign. (b) Uptime fraction versus time of day for the combined data including the fiber link (bars) and for the two clocks (lines). \label{fig:mask}}
\end{figure*}

The combined measurement uptime of 51\,\% (12.6 of 24.7 days) is distributed non-uniformly over the time of day (\autoref{fig:mask}). The daytime hours 06:00--18:00 contribute 43\,\% of the samples instead of 50\,\%. Uneven sampling could bias the mean only if the ratio itself varies with the time of day or the uptime of the clock. We find no significant deviation between the daytime and nighttime mean of $-2.0(1.9)\times10^{-18}$ and $+1.5(1.4)\times10^{-18}$ from the campaign mean within their uncertainties. Hourly means are consistent with a constant ($\chi^2_\mathrm{red} = 0.99$), and the data is uncorrelated with the hourly uptime ($r = 0.09$). These observations indicate that the non-uniform uptime distribution does not measurably affect the mean frequency ratio. 

As described in the main text, the statistical uncertainty for the measurement is derived under the assumption of white-frequency-noise scaling with $1.67\times10^{-15}/\sqrt{\tau(\mathrm{s})}$ observed in the Allan deviation for $\tau>1000$\,s. At 1000\,s averaging time, the data distribution is close to Gaussian, exhibiting much smaller, insignificant skewness and excess kurtosis of $\gamma_1 = -0.14(18)$ and $\gamma_2 = 1.0(7)$ (\autoref{fig:sim1000}(a)). Furthermore, simulations show that distributions with skewness and kurtosis matching either the 1000 s averages or even the unaveraged 1 s data lead to Allan deviations indistinguishable from those of Gaussian white-frequency noise, at all averaging times up to beyond $10^{7}$\,s (\autoref{fig:sim1000}(b)). These results support the use of a white-frequency-noise fit from 1000\,s onwards and its extrapolation to the full campaign duration.

\end{document}